\documentclass[pra,twocolumn,graphicx]{revtex4}
\usepackage{amssymb}
\usepackage{epsfig,amsmath}
\usepackage{graphicx}

\begin{document}

\title{Multielectron corrections in molecular high-order harmonic generation
for different formulations of the strong-field approximation}
\author{B.B. Augstein and C. Figueira de Morisson Faria}
\affiliation{Department of Physics and Astronomy, University College London, Gower
Street, London WC1E 6BT, United Kingdom}
\date{\today}

\begin{abstract}
We make a detailed assessment of which form of the dipole operator to use in
calculating high order harmonic generation within the framework of the
strong field approximation, and look specifically at the role the form plays
in the inclusion of multielectron effects perturbatively with regard to the
contributions of the highest occupied molecular orbital. We focus on how
these corrections affect the high-order harmonic spectra from aligned homonuclear
and heteronuclear molecules, exemplified by $\mathrm{N}_2$ and CO, respectively,
which are isoelectronic. We find that the velocity form
incorrectly finds zero static dipole moment in heteronuclear
molecules. In contrast, the length form of the dipole operator leads
to the physically expected non-vanishing expectation value for the
dipole operator in this case. Furthermore, the so called ``overlap''
integrals, in which the dipole matrix element is computed using
wavefunctions at different centers in the molecule, are prominent in
the first-order multielectron corrections for the velocity form, and
should not be ignored. Finally, inclusion of the multielectron
corrections has very little effect on the spectrum. This suggests that
relaxation, excitation and the dynamic motion of the core are
important in order to describe multielectron effects in molecular
high-order high harmonic generation.
\end{abstract}

\maketitle

\section{Introduction}

High harmonic generation can be used for attosecond imaging of molecules
\cite{Smirnova}. This is a consequence of the fact that the structure of an
aligned molecule can be inferred from quantum-interference patterns, which
appear in the high-harmonic spectra. Understood in terms of the three-step
model, in which an electron reaches the continuum by tunneling or
multiphoton ionization, is accelerated by the field and subsequently
recombines with a bound state of its parent molecule, emitting a high-order
harmonic photon \cite{Lewenstein}, this phenomenon occurs due to the
electron returning to different spatially separated centers. The simplest
scenario is high-harmonic generation in a diatomic molecule, for which, in
principle, recombination and thus high-harmonic emission can take place from
either atomic center. This is a microscopic equivalent to the double slit experiment \cite%
{doubleslit}.

How to accurately describe the whole process of high harmonic generation
(HHG) in molecules is still an open problem. There are two main theoretical
approaches that can be used. Firstly one can attempt a fully numerical
solution by using density functional theory, which gives accurate results
but inhibits a more complete understanding of the process \cite{tddft}.
Secondly, one can employ the traditional atomic three step model, within the
strong-field approximation \cite{Lewenstein}, and calculate the molecular
bound-state wavefunctions using a quantum chemistry code \cite{GAMESS}. In
this framework, the bound states of the molecule are mainly incorporated in
the ionization and recombination steps, in the form of prefactors. The latter
approach allows a better physical understanding of the problem. The price
one pays, however, is that, in order to be able to treat HHG within a
semi-analytical framework, several approximations must be made.

It is well known that the harmonic spectrum is proportional to the modulus
squared of the Fourier transform of the dipole acceleration. By using the
Ehrenfest theorem the dipole acceleration can be replaced by the dipole
velocity or the dipole length, with no loss of accuracy so long as the
integration is over all time and no physical approximations are made upon
all wavefunctions involved. However, when the SFA is used, this is no longer
the case, and replacement of the dipole operator gives rise to
inconsistencies in the SFA dipole matrix elements, which vary depending on
the form of the operator used. The most well known of these is the lack of
translational invariance when using the length form of the dipole operator.
In the single-active electron approximation, it has been shown that the
velocity form gives the most reliable results \cite{JMOCL2007} and is easier
to use than the acceleration form.

In calculating the SFA prefactors the simplest approach is to use the
highest occupied molecular orbital (HOMO) \cite{Milosevic}. However, the
problems that arise in this case are 1) which orbitals have the predominant
influence on high harmonic generation and 2) how to accurately include the
multielectron effects of electron exchange and correlation.

The first problem has to some extent been addressed in Ref. \cite{Smirnova}
by noting that the ionization potential of different orbitals varies with
the rotation of the molecule and hence the corresponding cutoffs in the high
harmonic spectra will be different. Furthermore, the symmetry of the orbital
will strongly affect the contribution of specific orbitals to high-order
harmonic spectra. This is particularly true as the alignment angle of the
molecule with respect to the laser-field polarization varies. In previous
work \cite{CarlaBrad} we also investigated high-order harmonic generation
beyond the single-active orbital approximation. In particular, we addressed
the quantum interference between the HOMO and the HOMO-1 of $\mathrm{N}_{2}$
and the HOMO and the lowest unoccupied molecular orbital (LUMO) of $\mathrm{N%
}_{2}^{+}$. For the latter case, we have also observed that electron
recombination to different orbitals can be mapped into different cutoffs in
the HHG spectra. Our investigations, however, did not incorporate electron
correlation. In fact, we have either dealt with simplified multielectron
models, in which the time evolution of the optically active electrons has
been decoupled, or with a coherent superposition of one-electron states.

The second problem is what we propose to investigate in this paper and in
doing so to consider the influence that these multielectron effects have on
high harmonic generation. In particular, we wish to incorporate such effects
perturbatively around the contributions of the highest occupied molecular
orbital. Although correlation in molecular high-order harmonic generation
has been addressed in Refs. \cite{Patchkovskii,Smirnova}, utilizing a
multielectron version of the strong-field approximation, such studies have
been performed numerically to a great extent. Including the correlation
analytically in the dipole matrix elements facilitates more insight into the
effects that correlation has on the high harmonic spectra. In Ref. \cite%
{Santra}, many-body perturbation theory was used to include electron
correlation as corrections to the standard three step model. Therein,
however, only the response of single atoms has been addressed.

From the viewpoint of which form of the dipole operator to use, we are
particularly interested in how such corrections are affected by the
different formulations, especially in molecular targets. Indeed, even in the
single-active electron approximation, the discrepant results obtained for
each dipole form have raised considerable debate. By extending the debate to
multielectron effects in molecules, more insight into which operator is most
accurate for harmonic generation, and indeed all strong field phenomena, is
gained. The use of molecules is beneficial because the two-center
interference pattern gives another indicator of how the formulation is
performing.

In order to make such an assessment, we will employ many-body perturbation
theory along the lines of Ref.~\cite{Santra} and the dipole moment in its
length and velocity forms. For that purpose, we will consider two
isoelectronic molecules: $\mathrm{N}_{2}$ and CO$.$ The HOMO, LUMO and
HOMO-1 in both molecules have similar geometry. The main difference between
them is that $\mathrm{N}_{2}$ is a homonuclear molecule, while CO is
heteronuclear. This means that, for the latter molecule, there is an
intrinsic static dipole moment. According to Ref.~\cite{Santra}, one expects
the multielectron corrections to be more prominent in this case.

This work is organized as follows. In Sec.~\ref{theory}, we provide the
theoretical framework which we will employ in order to compute the
high-order harmonic spectra. We then calculate the transition amplitudes, in
different formulations (Sec.~\ref{SFA}), within the strong field
approximation, and calculate the multielectron corrections (Sec.~\ref%
{corrections}), with particular focus on the form of the dipole operator.
Subsequently, in Sec.~\ref{results}, these expressions are employed to
compute the high-harmonic spectra. Finally, the paper is summarized in Sec.~%
\ref{concl}. Atomic units are used throughout.

\section{Theory}

\label{theory}

\subsection{Transition amplitude}

\label{SFA} The SFA transition amplitude in the particular formulation of
Ref. \cite{Lewenstein} reads

\begin{eqnarray}
b_{\Omega } &\hspace{-0.1cm}=\hspace*{-0.1cm}&i\int_{-\infty }^{\infty }%
\hspace*{-0.5cm}dt\int_{-\infty }^{t}~\hspace*{-0.5cm}dt^{\prime }\int
d^{3}ka_{\mathrm{rec}}^{\ast }(\mathbf{k}+\mathbf{A}(t))a_{\mathrm{ion}}(%
\mathbf{k}+\mathbf{A}(t^{\prime }))  \notag \\
&&\exp [iS(t,t^{\prime },\Omega ,\mathbf{k})]+c.c.,  \label{amplhhg}
\end{eqnarray}%
with the action
\begin{equation}
S(t,t^{\prime },\Omega ,\mathbf{k})=-\frac{1}{2}\int_{t^{\prime }}^{t}[%
\mathbf{k}+\mathbf{A}(\tau )]^{2}d\tau -\epsilon _{0}(t-t^{\prime })+\Omega t
\label{actionhhg}
\end{equation}%
and the prefactors $a_{\mathrm{rec}}(\mathbf{k}+\mathbf{A}(t))=\left\langle
\mathbf{k}+\mathbf{A}(t)\right\vert \mathbf{d}\cdot \mathbf{e}_{x}\left\vert
\Psi \right\rangle $ and $a_{\mathrm{ion}}(\mathbf{k}+\mathbf{A}(t^{\prime
}))=\left\langle \mathbf{k}+\mathbf{A}(t^{\prime })\right\vert \mathbf{E(}%
t^{\prime }\mathbf{)\cdot r}\left\vert \Psi \right\rangle .$ In Eqs. (\ref%
{amplhhg}) and (\ref{actionhhg}), $\mathbf{d}$, $\mathbf{e}_{x}$, $\epsilon
_{0},$ and $\Omega $ denote the dipole operator, the laser-polarization
vector, the ionization potential of the molecule in question, and the
harmonic frequency, respectively, and $\left\vert \Psi \right\rangle $ gives
the bound state with which the electron recombines, or from which it
tunnels. Eq.~(\ref{amplhhg}) is given in the length-gauge formulation of the
SFA. We will employ this gauge throughout. In the standard velocity-gauge
formulation for the molecular SFA, the two-center interference patterns
vanish \cite{gaugepapers,DM2009}. To first approximation, we assume that there is
one active electron, which tunnels from the highest occupied molecular
orbital, reaches the continuum and recombines. Many-body corrections due to
the presence of the other electrons will only be incorporated in the
prefactors.

The above-stated expression describes a physical process in which an
electron tunnels from its parent molecule at an instant $t^{\prime }$, and
propagates in the continuum from a time $t^{\prime }$ to a subsequent time $%
t,$ with the intermediate momentum $\mathbf{k}$. At $t,$ the electron
recombines with its parent ion, generating a high-energy photon of frequency
$\Omega .$ These steps are implicit in the action (\ref{actionhhg}). They
can also be directly identified in the saddle-point equations, obtained from
the values of $t,t^{\prime }$ and $\mathbf{k}$ which render the action
stationary. This implies $\partial S(t,t^{\prime },\Omega ,\mathbf{k}%
)/\partial t^{\prime }=\partial S(t,t^{\prime },\Omega ,\mathbf{k})/\partial
t=0$ and $\partial S(t,t^{\prime },\Omega ,\mathbf{k})/\partial \mathbf{k=0.}
$

These saddle-point equations read
\begin{equation}
\frac{\left[ \mathbf{k}+\mathbf{A}(t^{\prime })\right] ^{2}}{2}+\epsilon
_{0}=0,  \label{tunnelsame}
\end{equation}%
\begin{equation}
\int_{t^{\prime }}^{t}d\tau \left[ \mathbf{k}+\mathbf{A}(\tau )\right] =%
\mathbf{0}.  \label{return}
\end{equation}%
and
\begin{equation}
\frac{\left[ \mathbf{k}+\mathbf{A}(t)\right] ^{2}}{2}+\epsilon _{0}=\Omega .
\label{recsame}
\end{equation}%
Physically, Eq. (\ref{tunnelsame}) gives the kinetic energy of the electron
during tunneling. One should note that this equation has no real solution,
so that the tunneling time $t^{\prime }$ is complex. This is a consequence
of the fact that tunneling has no classical counterpart. Eq. (\ref{return})
constrains the momentum of the electron, so that it may return to the
geometrical center of its parent molecule. Finally, Eq. (\ref{recsame})
gives the conservation of energy upon recombination, in which the kinetic
energy of the returning electron is converted in a high-frequency photon.
The stationary-phase method will be employed to compute the spectra in this
work. For more details we refer to \cite{atiuni}.

\subsection{Many-body corrections}

\label{corrections}

We will now briefly discuss the many-body corrections derived in Ref. \cite%
{Santra}. They have been obtained by starting from a many-body dipole
operator in the recombination and ionization dipole matrix elements $a_{%
\mathrm{rec}}(\mathbf{k+A}(t))$ and $a_{\mathrm{ion}}(\mathbf{k+A}(t^{\prime
}))$, and applying M\"{o}ller Plesset \cite{Szabo} perturbation theory.

\begin{equation}
a_{\eta }(\mathbf{k})=a_{\eta }^{(0)}(\mathbf{k})+a_{\eta }^{(1)}(\mathbf{k}%
)+a_{\eta }^{(2)}(\mathbf{k})+O(H^{2}),  \label{main3}
\end{equation}%
where the dipole matrix element associated with the HOMO is given by
\begin{equation}
a_{\eta }^{(0)}(\mathbf{k})=\langle \mathbf{k}|\mathbf{d}|\psi _{0}\rangle .
\label{prefsinglee}
\end{equation}%
The first-order corrections read
\begin{equation}
a_{\eta }^{(1)}(\mathbf{k})=\sum_{i}(\mathbf{d}_{ii}\langle \mathbf{k}|\psi
_{0}\rangle -\mathbf{d}_{i0}\langle \mathbf{k}|\psi _{i}\rangle )
\label{corrections1}
\end{equation}%
and
\begin{eqnarray}
a_{\eta }^{(2)}(\mathbf{k}) &=&\sum_{i}\langle \mathbf{k}|\psi _{i}\rangle
\sum_{a}\sum_{j}\mathbf{d}_{aj}\frac{v_{a0[ij]}}{\epsilon _{i}+\epsilon
_{j}-\epsilon _{0}-\epsilon _{a}}+  \notag \\
&&\sum_{a}\langle \mathbf{k}|\psi _{a}\rangle \sum_{b}\sum_{i}\mathbf{d}_{ib}%
\frac{v_{ab[0j]}}{\epsilon _{0}+\epsilon _{i}-\epsilon _{a}-\epsilon _{b}},
\label{corrections2}
\end{eqnarray}%
where
\begin{equation}
\mathbf{d}_{\nu \mu }=\langle \psi _{\nu }|\mathbf{d}|\psi _{\mu }\rangle ,
\label{dipbound}
\end{equation}%
\begin{equation}
v_{a0[ij]}=\left\langle \psi _{i},\psi _{j}\right\vert v\left\vert \psi
_{a},\psi _{0}\right\rangle -\left\langle \psi _{j},\psi _{i}\right\vert
v\left\vert \psi _{a},\psi _{0}\right\rangle ,
\end{equation}%
and%
\begin{equation}
v_{ab[0j]}=\left\langle \psi _{0},\psi _{j}\right\vert v\left\vert \psi
_{a},\psi _{b}\right\rangle -\left\langle \psi _{j},\psi _{0}\right\vert
v\left\vert \psi _{a},\psi _{b}\right\rangle .
\end{equation}%
In the above-stated equations, $v$ denotes the electron-electron
correlation, $\epsilon _{\alpha }$ the orbital energies, and the index $\eta
$, with $\eta =$ (rec, ion), refers to recombination and ionization. The
indices $a,b$ represent unoccupied orbitals and the indices $i,j$ represent
occupied orbitals. The index 0 is related to the orbital around which the
corrections are inserted. In the specific scenario addressed in this paper,
this is the highest occupied molecular orbital (HOMO).

The indices $\nu ,\mu $ in the dipole matrix element (\ref{dipbound}) are
general. In the first-order corrections (\ref{corrections1}), they may relate to the expectation value of the dipole
operator ($\nu =\mu =i$), or to the dipole coupling between the HOMO and an
occupied bound state ($\nu =i$ and $\mu =0$). In the framework of the second-order corrections (\ref{corrections2}), they relate to the couplings
occupied and unoccupied bound states ($\nu =a$ and $\mu =j$ or $\nu =i$ and $\mu =b$).
Physically, $v_{a0[ij]}$ and $v_{ab[0j]}$ give the exchange terms, and are
dependent on the form of the electron-electron interaction. In the
corrections (\ref{corrections1}) and (\ref{corrections2}), the terms $%
\langle \mathbf{k}|\psi _{\nu }\rangle $ will determine the two-center
interference conditions, while the remaining terms will mainly act as
weights.

These corrections will also behave differently with regard to homonuclear
and heteronuclear molecules. For the former type of molecules, the orbitals
exhibit definite parities. Hence, the dipole matrix elements $\mathbf{d}%
_{\nu \mu }$ will only couple gerade and ungerade states. As \ a direct
consequence, the first term in (\ref{corrections1}) will vanish. Physically,
this is expected, as there is no static dipole moment for a homonuclear
molecule.

In contrast, for heteronuclear molecules, the matrix elements $\mathbf{d}%
_{\nu \mu }$ \ will in principle couple all states involved. The first term
in Eq. (\ref{corrections1}) will be non-vanishing, and, physically, is the
contribution of the static dipole moment of the molecule in question. This
term, however, will only lead to quantitative changes in the dipole matrix
element. In fact, an inspection of Eqs. (\ref{prefsinglee}) and (\ref%
{corrections1}) shows that it will yield the same two-center interference
condition as the one-particle prefactor $a_{\eta }^{(0)}(\mathbf{k})$.

In the above-mentioned framework, all the structure of the molecule is
embedded in the recombination prefactor $a_{\mathrm{rec}}(\mathbf{k}+\mathbf{%
A}(t)).$ This prefactor can be written in different forms, which will lead
to different results. Explicitly, the length, velocity and acceleration
forms of the dipole operator read
\begin{equation}
\mathbf{d}^{(l)}=\mathbf{\hat{r},}
\end{equation}%
\begin{equation}
\mathbf{d}^{(v)}=-\mathbf{\hat{k}}
\end{equation}%
and
\begin{equation}
\mathbf{d}^{(a)}=-\bigtriangledown {V(\mathbf{\hat{r}}),}
\end{equation}%
respectively, where the hats denote operators. The prefactor $a_{\mathrm{ion}%
}(\mathbf{k}+\mathbf{A}(t^{\prime }))$ will mainly influence the overall
harmonic intensity, as it is associated with ionization. In this work, we
will consider the length and the velocity forms of the dipole operator. In
the single-active-electron framework, the former and the latter lead to the
worst and best description of the two-center interference patterns \cite%
{JMOCL2007}.

\subsection{Molecular orbitals and interference condition}

In calculating the molecular orbitals we use a Linear Combination of Atomic
Orbitals (LCAO) approximation along with the Born Oppenheimer approximation.
In this framework, the molecular orbital wavefunction is
\begin{eqnarray}
\psi _{j}(\mathbf{r}) &=&\sum_{\alpha }c_{\alpha ,j}^{(L)}\Phi _{\alpha
,j}^{(L)}(\mathbf{r}+\mathbf{R}/2)  \notag \\
&&+(-1)^{l_{\alpha }-m_{\alpha }+\lambda _{\alpha }}c_{\alpha ,j}^{(R)}\Phi
_{\alpha ,j}^{(R)}(\mathbf{r}-\mathbf{R}/2)
\end{eqnarray}%
where \textbf{R} is the internuclear separation, $l_{\alpha }$ and $%
m_{\alpha }$ the orbital and magnetic quantum number respectively. The
coefficients $c_{\alpha ,j}^{(\xi )}$, with $\xi =L,R,$ form the linear
combination of atomic orbitals which are extracted from GAMESS-UK \cite%
{GAMESS}, and the indices $(L)$ and $(R)$ refer to the left or to the right
ion, respectively. The internuclear axis is taken to be in the $z$
direction, and the laser-field polarization is chosen along the radial
coordinate. For homonuclear molecules, $c_{\alpha ,j}^{(L)}=c_{\alpha
,j}^{(R)}=c_{\alpha ,j}$ and $\Phi _{\alpha ,j}^{(R)}=\Phi _{\alpha
,j}^{(L)}=\Phi _{\alpha ,j}.$ The parameter $\lambda _{\alpha }$ can be
varied depending on the symmetry. For homonuclear molecules, $\lambda
_{\alpha }$=$|m_{\alpha }|$ and $\lambda _{\alpha }$=$|m_{\alpha }|$+1 for
gerade and ungerade symmetry, respectively.

The wavefunctions themselves are then expanded as Gaussian type orbitals
with a 6-31 basis set, with
\begin{equation}  \label{contraction}
\Phi _{\alpha ,j}^{(\xi )}(\mathbf{r})=\sum_{\nu }b_{\nu ,j}^{(\xi
)}(r_{\chi })^{l_{\alpha }}e^{-\zeta _{\nu ,j}^{(\xi )}r^{2}}.
\end{equation}%
For the $\sigma ,$ $\pi _{x}$ and $\pi _{y}$ orbitals, $r_{\chi }=z,x$ and $%
y $, respectively. The index $\xi $ is related to the left or right ion, and
$b_{\nu ,j}^{(\xi )}$ and $\zeta _{\nu ,j}^{(\xi )}$ form the contraction
coefficients and exponents, also obtained from GAMESS-UK. Note that these
coefficients are real.

The momentum-space wavefunctions, which will be used in this work to compute
the prefactors, are given by%
\begin{eqnarray}
\psi _{j}(\mathbf{k}) &=&\sum_{\alpha }\exp [i\mathbf{k\cdot }\frac{\mathbf{R%
}}{2}]c_{\alpha ,j}^{(L)}\Phi _{\alpha ,j}^{(L)}(\mathbf{k}) \\
&&+(-1)^{l_{\alpha }-m_{\alpha }+\lambda _{\alpha }}\exp [-i\mathbf{k\cdot }%
\frac{\mathbf{R}}{2}]c_{\alpha ,j}^{(R)}\Phi _{\alpha ,j}^{(R)}(\mathbf{k}),
\notag  \label{orbitalpspace}
\end{eqnarray}%
where, similarly to the position-space situation%
\begin{equation}
\Phi _{\alpha ,j}^{(\xi )}(\mathbf{k})=\sum_{\nu }b_{\nu ,j}^{(\xi )}\tilde{%
\varphi}_{\nu ,j}^{(\xi )}(\mathbf{k}),
\end{equation}%
with
\begin{equation}
\tilde{\varphi}_{\nu ,j}^{(\xi )}(\mathbf{k})=(-ik_{\beta })^{l_{\alpha }}%
\frac{\pi ^{3/2}}{2^{l_{\alpha }}\left( \zeta _{\nu ,j}^{(\xi )}\right)
^{3/2+l_{\alpha }}}\exp [-k^{2}/(4\zeta _{\nu ,j}^{(\xi )})].
\end{equation}%
Therein, $\beta =z$, $\beta =x$ and $\beta =y$ for\ the $\sigma $, $\pi _{x}$
and $\pi _{y}$ orbitals, respectively. The return condition (\ref{return})
guarantees that the momentum $\mathbf{k}$ and the external field are
collinear. Hence, for a linearly polarized field $\theta _{k}$ is equal to
the alignment angle $\theta _{L}$ between the molecule and the field. The
above-stated equations have been derived under the assumption that only $s$
and $p$ states will be employed in order to build the wavefunctions employed
in this work. For more general expressions see Ref. \cite{CarlaBrad}.

\begin{figure}[tbp]
\includegraphics[width=9.5cm]{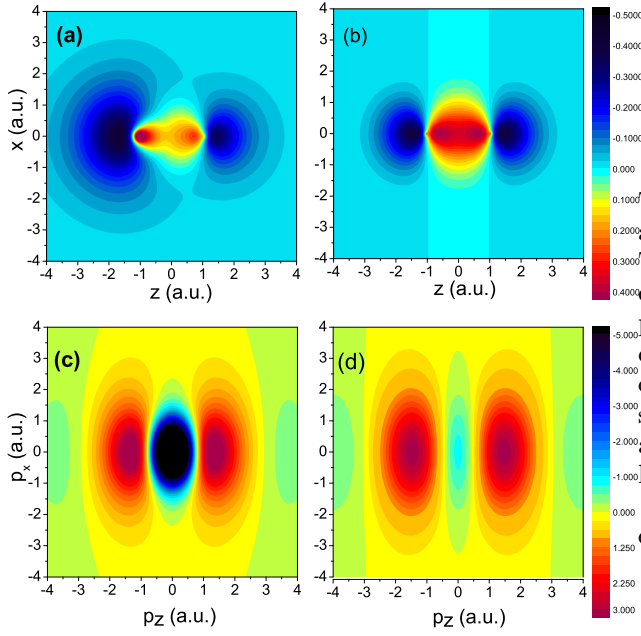}
\caption{Highest occupied molecular orbitals for CO [panels (a) and (c)] and
$\mathrm{N}_{2}$ [panels (b) and (b)], in the position and momentum space
(upper and lower panels, respectively).}
\label{fig1}
\end{figure}

In Fig.~1, we exhibit the HOMO for and CO and $\mathrm{N}_{2}$ obtained with
GAMESS-UK [Figs.~\ref{fig1}.(a) and (b), respectively]. The position-space
wavefunctions exhibit a central maxima and two minima located at the ion for
both species. The main difference is that, for CO, there is a strong bias
towards the Carbon atom, while for $\mathrm{N}_{2}$ the wavefunction is
symmetric. This is in agreement with the standard molecular-physics
literature. In contrast, we have found that both momentum-space
wavefunctions, given in Figs.~\ref{fig1}.(c) and (d), are symmetric with
respect to $(p_{x},p_{z})\rightarrow (-p_{x},-p_{z})$. In comparison to $%
\mathrm{N}_{2},$ CO exhibits a much deeper maximum at vanishing momenta.
Such maxima and minima may be determined by writing the exponents in Eq.~(%
\ref{orbitalpspace}) in terms of trigonometric functions. In this case, one
obtains%
\begin{equation}
\psi _{j}(\mathbf{k})=\sum_{\alpha }\mathcal{C}_{+}^{(\alpha )}\cos \left[
\frac{\mathbf{k\cdot R}}{2}\right] +i\mathcal{C}_{-}^{(\alpha )}\sin \left[
\frac{\mathbf{k\cdot R}}{2}\right] ,
\end{equation}%
with%
\begin{equation}
\mathcal{C}_{\pm }^{(\alpha )}=(-1)^{l_{\alpha }-m_{\alpha }+\lambda
_{\alpha }}c_{\alpha ,j}^{(R)}\Phi _{\alpha ,j}^{(R)}(\mathbf{k})\pm
c_{\alpha ,j}^{(L)}\Phi _{\alpha ,j}^{(L)}(\mathbf{k}).  \label{coeffinterf}
\end{equation}%
Calling $\vartheta =\arctan [i\mathcal{C}_{+}^{(\alpha )}/\mathcal{C}%
_{-}^{(\alpha )}],$we find
\begin{equation}
\psi _{j}(\mathbf{k})=\sum_{\alpha }\sqrt{\left( \mathcal{C}_{+}^{(\alpha
)}\right) ^{2}-\left( \mathcal{C}_{-}^{(\alpha )}\right) ^{2}}\sin
[\vartheta +\mathbf{k\cdot R}/2].  \label{pspaceinterf}
\end{equation}%
Eq.~(\ref{pspaceinterf}) exhibit minima for $\vartheta +\mathbf{k\cdot R}%
/2=n\pi$.

Note, however, that the coefficients $\mathcal{C}_{\pm }^{(\alpha )}$
defined in Eq.~(\ref{coeffinterf}) depend on the wavefunctions at the left
and right ions. Since these wavefunctions themselves depend on the momentum $%
\mathbf{k}$, one expects the two-center patterns to be blurred for
heteronuclear molecules. In contrast, for homonuclear molecules, $c_{\alpha
,j}^{(L)}=c_{\alpha ,j}^{(R)}$ and $\Phi _{\alpha ,j}^{(L)}(\mathbf{k})=\Phi
_{\alpha ,j}^{(R)}(\mathbf{k})$. This implies that the momentum dependence
in the argument $\vartheta$ cancels out and that the interference condition
in Refs.~\cite{CarlaBrad,Milosevic} is recovered. In this case, sharp
interference fringes are expected to be present.

The above-stated condition does not only lead to interference fringes in the
bound-state momentum wavefunctions but also in the high-harmonic spectra.
This is due to the fact that the dipole matrix elements depend on the
wavefunctions (\ref{orbitalpspace}). In this latter case, $\mathbf{k\rightarrow k+A}(t)$ in Eqs.~(\ref{orbitalpspace})-(\ref{pspaceinterf}).

\subsection{Dipole matrix elements}

In this section, we will provide explicit expressions for the dipole matrix
elements, depending on whether the dipole operator is given in its length or
velocity form. We will restrict ourselves to the first-order corrections to
the single-active-electron prefactor, as they are expected to be dominant.

\subsubsection{Single-active electron approximation}

We will commence with the single-active-electron dipole prefactor (\ref%
{prefsinglee}), which is employed in the standard strong-field
approximation. In the length form, this prefactor reads
\begin{equation}
a_{\mathrm{l}}^{(0)}(\mathbf{k})=-i\partial _{\mathbf{k}}\psi _{j}(\mathbf{k}%
),
\end{equation}%
where $\psi _{j}(\mathbf{k})$ is given by Eq. (\ref{orbitalpspace}). If
explicitly written in terms of the wavefunctions centered at the left and
right ions,

\begin{eqnarray}
a_{\mathrm{l}}^{(0)}(\mathbf{k}) &=&i\sum_{\alpha }c_{\alpha ,j}^{(L)}\exp [i%
\mathbf{k\cdot }\frac{\mathbf{R}}{2}]\partial _{\mathbf{k}}\Phi _{\alpha
,j}^{(L)}(\mathbf{k})  \notag \\
&&
+(-1)^{l_{\alpha }-m_{\alpha }+\lambda _{\alpha }}c_{\alpha ,j}^{(R)}\exp
[-i\mathbf{k}\cdot \frac{\mathbf{R}}{2}]\partial _{\mathbf{k}}\Phi _{\alpha ,j}^{(R)}(\mathbf{k})
\notag \\
&&+\frac{\mathbf{R}}{2}\phi _{j}(\mathbf{k}),
\label{alength}
\end{eqnarray}

with
\begin{eqnarray}
\phi _{j}(\mathbf{k}) &=&\sum_{\alpha }\exp [i\mathbf{k\cdot }\frac{\mathbf{R%
}}{2}]c_{\alpha ,j}^{(L)}\Phi _{\alpha ,j}^{(L)}(\mathbf{k}) \\
&&-(-1)^{l_{\alpha }-m_{\alpha }+\lambda _{\alpha }}c_{\alpha ,j}^{(R)}\exp
[-i\mathbf{k\cdot }\frac{\mathbf{R}}{2}]\Phi _{\alpha ,j}^{(R)}(\mathbf{k}).
\notag
\end{eqnarray}%
The first two terms in Eq. (\ref{alength}) \ are mainly oscillating, and
give the two-center interference maxima and minima. The third term in Eq. (%
\ref{alength}) increases with the internuclear distance, and has the main
effect of blurring such patterns. In fact, the existence of such term has
caused considerable debate and has been attributed to the lack of
orthogonality between the continuum and the bound-state wavefunctions in the
SFA \cite{dressedSFA}. Similar terms are also present in the first-order
corrections, to be discussed later.

In the velocity form, the single-electron dipole matrix element reads
\begin{eqnarray}
a_{\mathrm{v}}^{(0)} &=&\sum_{\alpha }c_{\alpha ,j}^{(L)}\mathbf{k}\exp [i%
\mathbf{k\cdot }\frac{\mathbf{R}}{2}]\Phi _{\alpha ,j}^{(L)}(\mathbf{k}) \\
&&+(-1)^{l_{\alpha }-m_{\alpha }+\lambda _{\alpha }}c_{\alpha ,j}^{(R)}%
\mathbf{k}\exp [-i\mathbf{k\cdot }\frac{\mathbf{R}}{2}]\Phi _{\alpha
,j}^{(R)}(\mathbf{k}).  \notag
\end{eqnarray}%
In contrast to the length-form scenario, there is no term linearly growing
with the internuclear separation. Indeed, previous studies for simplified
models for homonuclear molecules indicate that this is the preferred form of
the dipole operator, in order to obtain the correct interference patterns
\cite{JMOCL2007}.

\subsubsection{Corrections $a_{\protect\eta }^{(1)}(\mathbf{k})$}

We will now write down the explicit forms of Eq. (\ref{corrections1}) for
molecular orbitals in the LCAO approximation.\ According to Ref. \cite%
{Santra}, this term is responsible for the dominant corrections to the
single-electron prefactor, and depends on the momentum space wavefunction (%
\ref{orbitalpspace}), and on the matrix element (\ref{dipbound}). The former
is independent on the form of the dipole operator, while the latter is
explicitly given by the sum
\begin{equation}
\mathbf{d}_{\nu \mu }=\sum\limits_{\alpha ,l_{\alpha }}\sum\limits_{\beta
,l_{\beta }}\sum_{\xi ^{\prime },\xi }\Xi _{\xi ^{\prime },\xi }(l_{\alpha
},l_{\beta })\mathcal{I}_{\xi ^{\prime },\xi }^{\alpha ,\beta },
\label{dipolebound}
\end{equation}%
where%
\begin{eqnarray}
\mathcal{I}_{\xi ^{\prime },\xi }^{\alpha ,\beta } &=&\int d^{3}r\left[
c_{\alpha ,\nu }^{(\xi ^{\prime })}\right] ^{\ast }c_{\beta ,\mu }^{(\xi
)}\Phi _{\alpha ,\nu }^{\ast (\xi ^{\prime })}(\mathbf{r}+(-1)^{a}\mathbf{R}%
/2)  \notag \\
&&\times d(\mathbf{r})\Phi _{\beta ,\mu }^{(\xi )}(\mathbf{r}+(-1)^{b}%
\mathbf{R}/2).  \label{IntegralsLCAO}
\end{eqnarray}%
In Eq. (\ref{dipolebound}) and (\ref{IntegralsLCAO}), the indices $\xi
^{\prime },\xi $ are related to the right or left ions, the exponents $a,b$
are 1 and 0 for $\xi ^{\prime },\xi =R$ and $\xi ^{\prime },\xi =L,$
respectively, and $d(\mathbf{r})$ denotes the dipole operator in the
position representation. In its length and velocity forms, $d^{(l)}(\mathbf{r})=\mathbf{r}$ and $d^{(v)}(\mathbf{r})=i\nabla _{\mathbf{r%
}}$.
The coefficients $\Xi _{\xi ^{\prime },\xi }(l_{\alpha },l_{\beta })$ are
such that $\Xi _{L,L}(l_{\alpha },l_{\beta })=1,\Xi _{L,R}(l_{\alpha
},l_{\beta })=(-1)^{l_{\beta }+\lambda _{\beta }},\Xi _{R,L}(l_{\alpha
},l_{\beta })=(-1)^{l_{\alpha }+\lambda _{\alpha }}$ and $\Xi
_{R,R}(l_{\alpha },l_{\beta })=(-1)^{l_{\beta }+\lambda _{\beta }+l_{\alpha
}+\lambda \alpha }.$

In Eq.~(\ref{IntegralsLCAO}), one may recognize two types of contributions.
The direct integrals
\begin{equation}
\mathcal{I}_{\xi ,\xi }^{\alpha ,\beta }=\int d^{3}r\left[ c_{\alpha ,\nu
}^{(\xi )}\right] ^{\ast }c_{\beta ,\mu }^{(\xi )}\Phi _{\alpha ,\nu }^{\ast
(\xi )}(\mathbf{r}\pm \mathbf{R}/2)d(\mathbf{r})\Phi _{\beta ,\mu }^{(\xi )}(%
\mathbf{r}\pm \mathbf{R}/2),  \label{direct}
\end{equation}%
which contains one-electron wavefunctions centered at the same ion, and the
so-called \textquotedblleft overlap" integrals
\begin{equation}
\mathcal{I}_{\xi ^{\prime },\xi }^{\alpha ,\beta }=\int d^{3}r\left[
c_{\alpha ,\nu }^{(\xi ^{\prime })}\right] ^{\ast }c_{\beta ,\mu }^{(\xi
)}\Phi _{\alpha ,\nu }^{\ast (\xi ^{\prime })}(\mathbf{r}\pm \mathbf{R}/2)d(%
\mathbf{r})\Phi _{\beta ,\mu }^{(\xi )}(\mathbf{r}\mp \mathbf{R}/2),
\label{overlap}
\end{equation}%
for $\xi ^{\prime }\neq \xi ,$ with one-electron wavefunctions localized at
different ions. For the moment ignoring the overlap integrals,
\begin{eqnarray}
\mathbf{d}_{\nu \mu } &=&\sum\limits_{\alpha ,l_{\alpha }}\sum\limits_{\beta
,l_{\beta }}\left[ c_{\alpha ,\nu }^{(L)}\right] ^{\ast }c_{\beta ,\mu
}^{(L)}\mathcal{I}_{L,L}^{\alpha ,\beta }  \label{dipmolecule} \\
&&+(-1)^{l_{\beta }+\lambda _{\beta }+l_{\alpha }+\lambda _{\alpha }}\left[
c_{\alpha ,\nu }^{(R)}\right] ^{\ast }c_{\beta ,\mu }^{(R)}\mathcal{I}%
_{R,R}^{\alpha ,\beta }.  \notag
\end{eqnarray}%
The explicit expressions for the overlap integrals are provided in Appendix
1.

We will commence by providing expressions for (\ref{dipolebound}) using the
velocity form. For that purpose, one must solve the integrals (\ref%
{IntegralsLCAO}). Depending on whether the dipole operator couples only $%
\sigma $ or $\pi $ orbitals, or a $\sigma $ orbital with a $\pi $ orbital,
these integrals will be different. If only $\sigma $ orbitals are involved,%
\begin{eqnarray}
\mathcal{I}_{\xi ,\xi }^{\alpha ,\beta }(\sigma ,\sigma )\hspace*{-0.2cm} &=&%
\hspace*{-0.25cm}\sum_{j,j^{\prime }}\frac{\pi b_{j,\nu }^{(\xi
)}b_{j^{\prime },\mu }^{(\xi )}}{(\zeta _{j,\nu }^{(\xi )}+\zeta _{j^{\prime
},\mu }^{(\xi )})}  \label{velocitysigmasigma} \\
&&\hspace*{-0.2cm}\times \left\{ l_{\beta }\mathcal{F}(l_{\beta }+l_{\alpha
}-1)-2\zeta _{j^{\prime },\mu }^{(\xi )}\mathcal{F}(l_{\beta }+l_{\alpha
}+1)\right\} ,  \notag
\end{eqnarray}%
with $\xi =R$ or $L,$ and%
\begin{equation}
\mathcal{F}(l)=\frac{1}{2}\left[ 1+(-1)^{l}\right] (\zeta _{j,\nu }^{(\xi
)}+\zeta _{j^{\prime },\mu }^{(\xi ^{\prime })})^{-1/2-l/2}\Gamma \left[
\frac{1+l}{2}\right] .  \label{F}
\end{equation}%
The same expression is encountered if only $\pi _{x}$ or $\pi _{y}$ are
coupled, i.e., $\mathcal{I}_{\xi ,\xi }^{\alpha ,\beta }(\sigma ,\sigma )=%
\mathcal{I}_{\xi ,\xi }^{\alpha ,\beta }(\pi _{x},\pi _{x})=\mathcal{I}_{\xi
,\xi }^{\alpha ,\beta }(\pi _{y},\pi _{y}).$ The above-stated expression is
only non-vanishing if $l_{\beta }+l_{\alpha }+1$ is even. In our framework,
this implies that only $s$ and $p$ states are mixed. \ If, on the other
hand, a $\sigma $ orbital is coupled to a $\pi _{x}$ or a $\pi _{y}$
orbital, this integral reads
\begin{eqnarray}
\mathcal{I}_{\xi ,\xi }^{\alpha ,\beta }(\sigma ,\pi _{\chi })
&=&\sum_{j,j^{\prime }}\frac{\sqrt{\pi }b_{j,\nu }^{(\xi )}b_{j^{\prime
},\mu }^{(\xi )}}{\left( \zeta _{j,\nu }^{(\xi )}+\zeta _{j^{\prime },\mu
}^{(\xi )}\right) ^{1/2}}  \notag \\
&&\times (\mathcal{B}(l_{\alpha },l_{\beta })-2\zeta _{j^{\prime },\mu
}^{(\xi )}\mathcal{A}(l_{\alpha }+1,l_{\beta })),  \label{integralsigmpi}
\end{eqnarray}%
with $\chi =x,y.$ Thereby,
\begin{equation}
\mathcal{A}(l_{\alpha },l_{\beta })=\mathcal{F}(l_{\alpha })\mathcal{F}%
(l_{\beta }),  \label{pisigm1}
\end{equation}%
\begin{equation}
\mathcal{B}(l_{\alpha },l_{\beta })=\mathcal{F}(l_{\alpha })\left\{ l_{\beta
}\mathcal{F}(l_{\beta }-1)-2\zeta _{j^{\prime },\mu }^{(\xi )}\mathcal{F}%
(l_{\beta }+1)\right\} ,  \label{pisigm2}
\end{equation}%
and $\mathcal{F}(l_{\beta })$ is defined according to Eq. (\ref{F}). Eq.~(%
\ref{integralsigmpi}) also holds for the dipole matric elements coupling the
$\pi _{x}$ and $\pi _{y}$ orbitals, i..e, $\mathcal{I}_{\xi ,\xi }^{\alpha
,\beta }(\sigma ,\pi _{\chi })=\mathcal{I}_{\xi ,\xi }^{\alpha ,\beta }(\pi
_{x},\pi _{y})=\mathcal{I}_{\xi ,\xi }^{\alpha ,\beta }(\pi _{y},\pi _{x}).$
From Eqs.~(\ref{pisigm1}) and (\ref{pisigm2}) we see that $\mathcal{A}%
(l_{\alpha }+1,l_{\beta })\neq 0$ if $l_{\alpha }=0$ and $l_{\beta }=1,$ and
that $\mathcal{B}(l_{\alpha },l_{\beta })\neq 0$ $\ $if $l_{\alpha }=1$ and $%
l_{\beta }=0.$ Hence, once more, only $s$ and $p$ states are coupled.

All these contributions are then included in the corrections (\ref%
{corrections1}). In the specific case of a homonuclear molecule, an
inspection of Eqs.~(\ref{dipmolecule})-(\ref{pisigm2}) shows that only
orbitals of different parities are coupled. In fact, if $c_{\alpha ,\nu
}^{(L)}=c_{\alpha ,\nu }^{(R)}$ and $c_{\beta ,\mu }^{(L)}=c_{\beta ,\mu
}^{(R)}$, Eq.~(\ref{dipmolecule}) is only nonvanishing if $l_{\beta
}+\lambda _{\beta }+l_{\alpha }+\lambda _{\alpha }$ is even. Since, as
discussed above, $\mathcal{I}_{\xi ,\xi }^{\alpha ,\beta }$ are only
nonvanishing if $l_{\alpha }$ and $l_{\beta }$ have values corresponding to different parities, this
must also hold for $\lambda _{\beta }$ and $\lambda _{\alpha }$. In contrast, for heteronuclear molecules, these
coefficients are different and in principle all terms contribute.

We will now state the length-form dipole matrix element. We will mainly
consider the integrals $\mathcal{I}_{\xi ,\xi }^{\alpha ,\beta }.$ If only $%
\sigma $ orbitals are coupled, these integrals read%
\begin{eqnarray}
\mathcal{I}_{\xi ,\xi }^{\alpha ,\beta }(\sigma ,\sigma ) &=&\hspace*{-0.25cm%
}\sum_{j,j^{\prime }}\frac{\pi b_{j,\nu }^{(\xi )}b_{j^{\prime },\mu }^{(\xi
)}}{(\zeta _{j,\nu }^{(\xi )}+\zeta _{j^{\prime },\mu }^{(\xi )})} \\
&\times&\left[ \mathcal{F}(l_{\alpha }+l_{\beta }+1)\mp \frac{R}{2}\mathcal{F%
}(l_{\alpha }+l_{\beta })\right] ,  \notag  \label{lengthsigmsigm}
\end{eqnarray}%
The same expression holds for transitions involving only the $\pi _{x}$ or
the $\pi _{y}$ orbitals, i.e., $\mathcal{I}_{\xi ,\xi }^{\alpha ,\beta
}(\sigma ,\sigma )=\mathcal{I}_{\xi ,\xi }^{\alpha ,\beta }(\pi _{x},\pi
_{x})=\mathcal{I}_{\xi ,\xi }^{\alpha ,\beta }(\pi _{y},\pi _{y})$.

If the dipole operator couples $\sigma $ with $\pi $ orbitals, such
integrals are given by
\begin{eqnarray}
\mathcal{I}_{\xi ,\xi }^{\alpha ,\beta }(\sigma ,\pi _{\chi })
&=&\sum_{j,j^{\prime }}\frac{\sqrt{\pi }b_{j,\nu }^{(\xi )}b_{j^{\prime
},\mu }^{(\xi )}}{(\zeta _{j,\nu }^{(\xi )}+\zeta _{j^{\prime },\mu }^{(\xi
)})^{1/2}} \\
&\times &\left[ \mathcal{A}(l_{\alpha }+1,l_{\beta })+\mathcal{A}(l_{\alpha
},l_{\beta }+1)\mp \frac{R}{2}\mathcal{A}(l_{\alpha },l_{\beta })\right] ,
\notag  \label{lengthsigmpi}
\end{eqnarray}%
with $\chi =x,y.$ Eq. (\ref{lengthsigmpi}) also holds for couplings between
different $\pi $ orbitals, i.e., $\mathcal{I}_{\xi ,\xi }^{\alpha ,\beta
}(\sigma ,\pi _{\chi })=\mathcal{I}_{\xi ,\xi }^{\alpha ,\beta }(\pi
_{y},\pi _{x})=\mathcal{I}_{\xi ,\xi }^{\alpha ,\beta }(\pi _{x},\pi _{y}).$

In the above-stated expressions, one may identify two distinct behaviors.
The terms $\mathcal{F}(l_{\alpha }+l_{\beta }+1),$ $\mathcal{A}(l_{\alpha
}+1,l_{\beta })$ and $\mathcal{A}(l_{\alpha },l_{\beta }+1)$ are
non-vanishing only if $l_{\alpha }$ and $l_{\beta }$ have values
corresponding to different parities, i.e., in our framework, it
couples $s$ and $p$ states. The terms $\mathcal{F}(l_{\alpha }+l_{\beta })$ and $\mathcal{A}(l_{\alpha },l_{\beta
}),$ on the other hand, couple states of the same parity (either $s$ or $p$%
). For homonuclear molecules, these terms will vanish, as the dipole
operator will only couple states of different parities, i.e., $l_{\beta
}+l_{\alpha }+1$ must be even in Eq. (\ref{dipmolecule}). For heteronuclear
molecules, in contrast, they may in principle be present as the coefficients
at each center of the molecule are different.

\subsection{Inconsistencies in the dipole matrix elements}

We will now discuss some of the subtleties that arise in the evaluation of
our multielectron corrections. Firstly, the overlap integrals, in Eq. (\ref%
{overlap}) are often presumed to be very small in comparison to their direct
counterpart, but this is not always the case. This can be seen by using the
Gaussian theorem,
\begin{equation}
e^{-\zeta_{\alpha}(r-\mathbf{R}/2)^2}e^{-\zeta_{\beta}(r+\mathbf{R}%
/2)^2}=Ke^{-\rho(r-\mathbf{R}_p)^2}
\end{equation}
where
\begin{equation}
\rho=\zeta_{\alpha}+\zeta_{\beta},
\end{equation}
\begin{equation}
K=e^{-\frac{\zeta_\alpha\zeta_\beta}{\zeta_\alpha+\zeta_\beta}|\mathbf{R}%
|^2},  \label{gaussianoverl}
\end{equation}
and
\begin{equation}
\mathbf{R}_p=\frac{\mathbf{R}(\zeta_{\alpha} - \zeta_{\beta})}{%
2(\zeta_{\alpha}+\zeta_{\beta})}.
\end{equation}
Here $\zeta_{\alpha}$ and $\zeta_{\beta}$ correspond to our exponents in Eq.
(\ref{contraction}).

At first sight it would appear that these contributions are much smaller
than those from the direct integrals (\ref{direct}). This does not hold,
however, for exponents that are less than unity, of which there are many for
various different atomic orbitals. It is therefore not astute to ignore the
overlap integrals in this instance.

In fact, we have verified that in most cases the overlap integrals (\ref%
{overlap}) are comparable to the integrals (\ref{direct}) at the same
center. More importantly, we have also found that the velocity form gives zero static dipole moment for
heteronuclear molecules when using a Gaussian basis set. This is an unphysical
result, which has been verified for the direct integrals by using the properties of Gamma functions
to analyze the term in brackets in Eq.~(\ref{velocitysigmasigma}).
Explicitly, it can be shown that this term is proportional to $[\zeta
_{j}l_{\beta }-\zeta _{j}^{\prime }l_{\alpha }/(\zeta _{j}^{\prime }+\zeta
_{j})]$. Since only $s$ and $p$ states can be coupled, each pair $(l_{\alpha},l_{%
\beta})=(0,1)$ and $(l_{\alpha},l_{\beta})=(1,0)$ will lead to
symmetric contributions which will cancel out. We have also verified numerically that
the overlap integrals are vanishingly small in this case.

Apart from that, inclusion of the overlap integrals gives rise to
terms that depend on R, and thus removes the translational
invariance from the velocity form. This problem is also encountered
in the single-active electron matrix element when using the length
form. In contrast to the one-electron length-form scenario, however,
these terms will vanish at large internuclear distances due to the
Gaussian exponentials (\ref{gaussianoverl}). (see a discussion in
\cite{dressedSFA} for their counterpart in the single-active
electron framework)

In contrast, the length form predicts non-zero static dipole moment
in heteronuclear molecules, which makes sense physically. However,
for the heteronuclear case the R-dependent terms that arise in the
direct integrals couple states of
the same parity as seen in Eq.~(\ref{lengthsigmsigm}) and Eq.~(\ref%
{lengthsigmpi}). This should not occur for an odd operator such as the
dipole, which is an unphysical result. This is another reason to suggest that
these terms should be neglected. In the results that follows, this term will be neglected unless otherwise stated.

\section{High-order harmonic spectra}

\label{results}

Below we compare the harmonic spectra predicted by different forms of the
dipole operator. We will place particular emphasis on the comparison between
homonuclear and heteronuclear molecules. The former and the latter will be
exemplified by $\mathrm{N}_{2}$ and CO, respectively. We will consider the
full, three-dimensional dynamics of the problem. For that purpose, one must
integrate over the azimuthal coordinate, $\phi _{p}$, such that,
\begin{equation}
S(\Omega )=\int_{0}^{2\pi }|M|^{2}d\phi _{k},
\end{equation}%
where $\ M$ is the overall transition amplitude. This allows one to consider
the degeneracy of the $\pi $ orbitals \cite{CarlaBrad}. The corrections are
incorporated in the ionization and recombination prefactors. The former gives the
particular weight for tunneling to take place, and the latter gives the shape of the
high-order harmonic spectra and the two-center fringes. The explicit
expressions for the dipole matrix elements are provided in Appendix 2.

\begin{figure}[tbp]
\includegraphics[width=9.5cm]{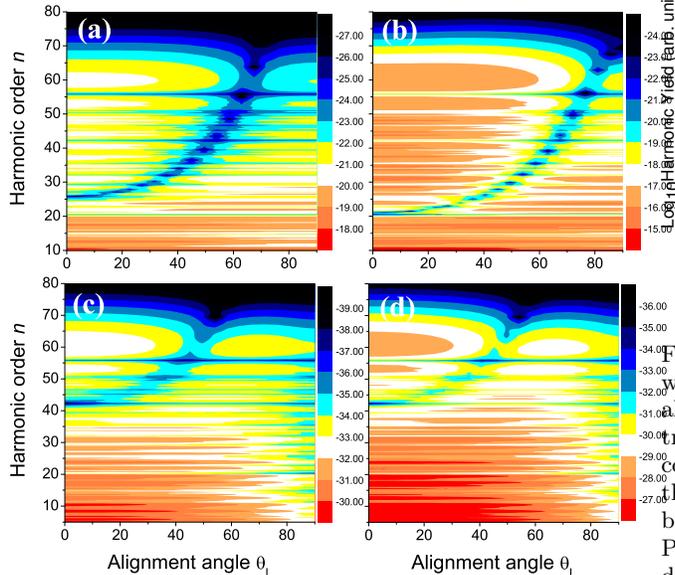}
\caption{High-order harmonic spectra for $\mathrm{N}_2$ subject to a
linearly polarized laser field of frequency $\protect\omega=0.057$
a.u. and intensity $I=4 \times 10^{14}\mathrm{W/cm}^2$, as a
function of the alignment angle $\protect\theta_L$ between the
molecule and the field. In the figure we consider the corrections in
both the ionization and the recombination prefactors. Panels (a) and
(c) have been computed employing the length form, while panels (b)
and (d) have been calculated using the velocity form. The upper
panels display the $3\protect\sigma_g$ spectrum in which the
first-order corrections have been incorporated. In the lower panels,
we depict the yield obtained with the first order corrections in the
length and the velocity form. The bound-state energy of the HOMO and
the
equilibrium internuclear distance have been taken from GAMESS-UK. Explicitly, $%
\protect\epsilon^{(\mathrm{N}_2)}_{0}=0.63485797$ a.u and $R^{(\mathrm{N}%
_2)}=2.068$ a.u. The contour plots have been normalized with regard
to the maximum yield in each panel.} \label{N2ionrec}
\end{figure}
We will now start by discussing the high-harmonic spectra of
diatomic nitrogen ($\mathrm{N}_2$) computed including the
single-atom prefactor and the first-order corrections. In
Fig.~\ref{N2ionrec}, we display such spectra, in the velocity and
length form (panels (a) and (b), respectively), together with the
corrections alone (panels (c) and (d), respectively). In the
length-form prefactor $a_{\sigma }^{(0)}(\mathbf{k})$, we omitted
the term growing linearly with the internuclear distance $R$. In the
corrections, we included both the direct and the overlap integrals.

In the figure, we notice a clear two-center minimum, which is due to
the interference of high-harmonic emission at the two spatially
separated centers in the molecule. This minimum varies with the
alignment angle $\theta_L$ of the molecule with respect to the
laser-field polarization. In the upper panels, its energy position
is characteristic of the $3\sigma_g$ orbital
\cite{DM2009,CarlaBrad}. A comparison of panels (a) and (b) shows
that the two-center minimum is slightly displaced, depending on the
form of the dipole operator. For instance, for vanishing alignment
angle, in the velocity form this minimum is near $\Omega=21\omega$,
while in the length form it is at a slightly higher energy
($\Omega=25\omega$). This is due to the fact that the $s-p$ mixing,
which determines the energy position of this minimum, is different
for each case. The velocity form slightly favors the contributions
from the $p$ states, compared to the length form. For a discussion
of the $s-p$ mixing in a slightly different context, namely of
different basis sets, and its influence on the high-order harmonic
spectra, see \cite{CarlaBrad}. Here, however, this bias is
introduced by the different forms of the dipole operator.

The remaining panels exhibit the spectra obtained employing the
first-order corrections only. Such contributions exhibit a
well-defined minimum, whose position is at a quite different energy
from that observed in the overall prefactor. We have verified that
this pattern can be related to the contributions of the $\sigma_u$
orbitals. The contributions of the $\pi_u$ orbitals are strongly
suppressed in this parameter region due to to the presence of a
nodal plane along the internuclear axis.

Quantitatively, the yield in panels (c)  and (d) is several orders
of magnitude smaller than those in the upper panels. This shows
that, in the present framework, it suffices to consider the zeroth
order contributions. This behavior occurs both for the length and
the velocity form of the dipole operator, and is due to the fact
that the corrections are considerably smaller both in the ionization
and recombination prefactor.  The above-mentioned suppression of
ionization for the $\pi$ orbitals also contributes to such results.

\begin{figure}[tbp]
\includegraphics[width=9cm]{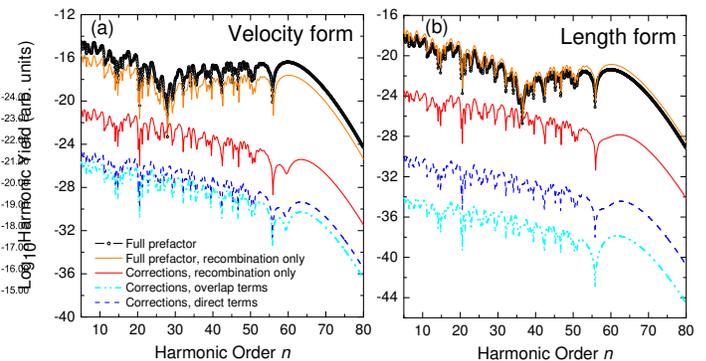}
\caption{High-order harmonic spectra for $\mathrm{N}_2$ in a driving
field with the same parameters as in the previous figure, and an
alignment angle $\theta_L=\pi/4$. The figure shows the overall
spectra together with the first-order corrections taking into
account only the direct or overlap integrals. We consider both the
ionization and recombination prefactors, or the recombination
prefactor only (orange and red lines in the figure). Panels (a) and
(b) give the velocity and length forms of the dipole operator,
respectively.} \label{cutsover}
\end{figure}

The influence of the ionization prefactor can be seen in
Fig.~\ref{cutsover}. Therein, we present spectra computed for a
fixed alignment angle assuming
$a_{\mathrm{ion}}(\mathbf{k}+\mathbf{A}(t_0))$ to be constant and
equal to unity. Physically, this implies that any influence of the
orbital symmetry on tunneling ionization, such as the presence or
absence of nodal planes, has been removed, both for the zeroth-order
prefactor and the corrections. For comparison, we are also providing
the full spectra, with both the ionization and recombination
prefactors.

The figure shows that, while the overall spectra changes relatively
little, or may even increase if this prefactor is removed (see
orange and black lines in Fig.~\ref{cutsover}.(a)), the corrections
increase in several orders of magnitude. This can be attributed to
the following. In the parameter range of interest, the zeroth order
ionization prefactor, which is the main influence in the overall
spectra, is of the order of unity or even slightly larger for a
$3\sigma_g$ orbital. This is due to the absence of nodal planes for
either parallel or perpendicular alignment in this particular case.
In contrast, in the corrections, mainly $\sigma_u$ or $\pi_u$
orbitals are involved. The former orbitals exhibit nodal planes for
alignment angles $\theta_L=90^{\circ}$ and the latter for
$\theta_L=0$. This will cause an overall suppression in tunnel
ionization for a wide range of angles, when such orbitals are summed
over. Furthermore, the dipole matrix elements $d_{\mu\nu}$ in the
spectra are quite small and will also contribute to the
above-mentioned suppression.

Another noteworthy feature also presented in Fig.~\ref{cutsover} is
the influence of the so-called overlap integrals, in which the
dipole operator couples wavefunctions at different centers in the
molecule, in the multielectron corrections. In general, these
integrals are assumed to be very small in comparison with the direct
integrals, in which wavefunctions at the same ion are taken. Hence,
in many cases the overlap integrals are neglected. We have verified,
however, that this depends strongly on the form of the dipole
operator.
 For the velocity-form dipole, the
spectra obtained from the overlap integrals are comparable to those
obtained from the direct integrals only [Fig.~\ref{cutsover}.(a)].
Hence, both contributions must be included. In contrast, if the
length form is considered, the contributions from the overlap
integrals are considerably smaller than their direct-integral
counterparts [Fig.~\ref{cutsover}.(b)]. Therefore, the former may be
neglected.
\begin{figure}[tbp]
\includegraphics[width=9cm]{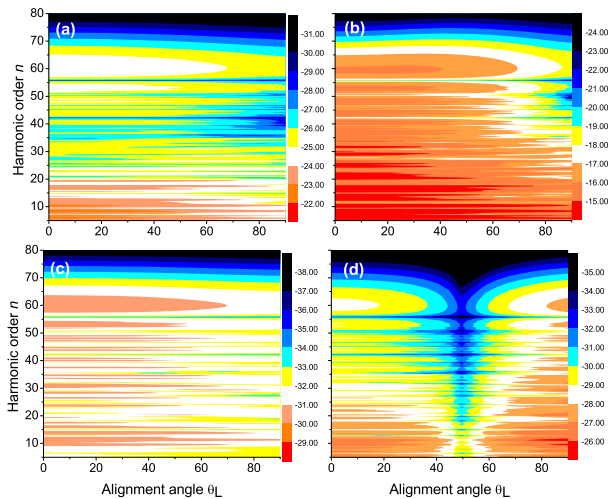}
\caption{High-order harmonic spectra for CO in a driving field with
the same parameters as in the previous figure, as a function of the alignment angle. Panels (a) and (c) exhibit the full spectra and the corrections in the length form, while panels (b) and (d) give their counterparts in the velocity form. In the length form, the linearly growing terms with regard to the internuclear distance have been neglected. } \label{figCO}
\end{figure}

Next, we will discuss the spectra encountered for CO. The overall
spectra are depicted in Fig.~\ref{figCO}, together with the
first-order corrections (upper and lower panels, respectively). As
an overall feature, the spectra exhibit no perceivable two-center
interference minimum, as CO is a heteronuclear molecule. This means that, upon
recombination, there will be more high-order harmonic emission from
a specific spatial region, where the orbital with which the electron
recombines, i.e., the HOMO, is more localized. This occurs both for the velocity and
length forms (Figs.~\ref{figCO}.(a) and (b), respectively).

 One notices, however,
that the decrease in the overall yield as the molecular alignment angle
approaches $\theta_L=90^{\circ}$ remains, due to the fact that the
HOMO is a $\sigma$ orbital, and thus highly localized along the
internuclear axis. As the laser-field polarization direction moves
away from this axis, recombination, and hence high-order harmonic
emission becomes less and less probable. Due, however, to the heteronuclear nature of this molecule, this suppression occurs at a different angle compared to $\mathrm{N}_2$.

Once more, we find that the contributions of the first-order
multielectron corrections to the spectra are several orders of
magnitude smaller than the zeroth-order terms. These contributions
are depicted in Fig.~\ref{figCO}.(c) and (d), for the length and velocity form, respectively. For such corrections, the interference minimum due to
high-harmonic emission at different centers has been washed out. This minimum was very clear
for $\mathrm{N}_2$ (see
Figs.~\ref{N2ionrec}.(c) and (d) for comparison). There is, however,
a very pronounced minimum near $\theta_L=50^{\circ}$ for the velocity-form corrections, which is independent of the harmonic frequency (see Fig.~\ref{figCO}.(d)). This minimum is
due to the geometry of the three lower-lying $\sigma$ orbitals, which influence the
corrections. This suggests that the imprints on the high-order
harmonic spectra due to purely geometric features, such as nodal
planes or minima in the orbitals, are more robust than those related
to quantum interference between harmonics emitted as spatially
separated centers. The length-form corrections, in contrast, do not exhibit any such feature. In fact, we have verified that they are dominated by the static dipole moment, which behaves in the same way as the HOMO.

\begin{figure}[tbp]
\includegraphics[width=9.5cm]{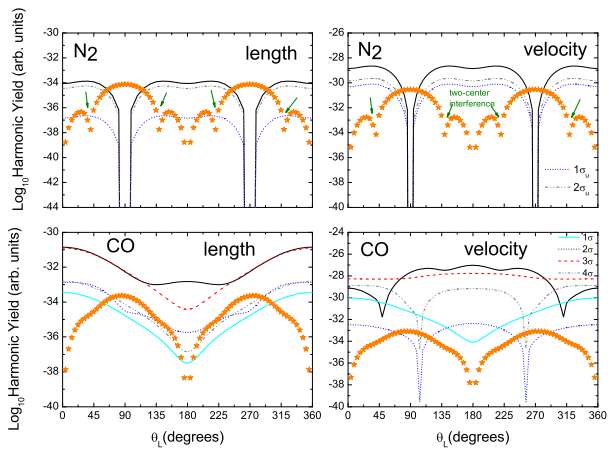}
\caption{Contributions of individual orbitals to the first-order corrections in the $45_{th}$ harmonic of a field with the same parameters as in the previous figures, as functions of the alignment angle $\theta_L$ of the molecule relative to the laser-field polarization. The thick black lines in the panel give the overall contributions of all $\sigma$ orbitals and the orange stars the contributions of the $\pi$ orbitals, respectively. Panels (a) and (b) refer to $\mathrm{N}_2$, and panels (c) and (d) correspond to CO, respectively.  The left and right panels display the results using the length and velocity form of the dipole operator, respectively. The minima due to the two-center interference are indicated by green arrows in the figure. } \label{cutsCO}
\end{figure}

Below, in Fig.~\ref{cutsCO}, we perform a more detailed analysis of
the above-mentioned issue. For that purpose, we consider the
contributions of individual orbitals to the corrections, and their
influence on the harmonic spectra, for fixed harmonic order. We
consider a mid-plateau harmonic, namely $\Omega=45\omega$. These
results are displayed for both  $\mathrm{N}_2$ and CO, as functions
of the alignment angle (upper and lower panels in the figure,
respectively). As an overall feature, we observe that, for a
homonuclear molecule, all contributions are symmetric upon $\theta
\rightarrow \theta + \pi$ (see panels(a) and (b)). Physically, this is expected, as such
molecules exhibit inversion symmetry and a rotation of $\pi$ should
not alter the physics of the problem. Clearly, this no longer holds
for a heteronuclear molecule. In this latter case,
however, the contribution of $S_n(\theta_L)$ to the yield is the mirror
image of $S_n(\theta_L+\pi)$. This is clear, as in this case the
right and the left ions are reversed.

The individual contributions provide useful information about the
geometry of the orbitals involved in the corrections. This can be
seen in all cases, regardless of whether the molecule is
heteronuclear or homonuclear. For instance, the contributions from
the $\pi$ orbitals, shown as the orange symbols in the figure, decay
very abruptly for $\theta_L=n\pi$. This is due to the fact that such
orbitals exhibit nodal planes in the case of $\mathrm{N}_2$ and a feature similar to nodal planes in the case of CO. The main difference between the homonuclear and the heteronuclear case is the two-center interference minimum, which is absent in the heteronuclear case. Furthermore, the shapes of such distributions do not change with the form of the dipole operator.  Any changes are mostly quantitative (for a comparison, see panels (a) and (b) for $\mathrm{N}_2$ and panels (c) and (d) for CO).

The contributions from the $\sigma$ orbitals, on the other hand, can behave in very different ways depending on whether the velocity or the length form is taken, in the heteronuclear case, such as CO. For $\mathrm{N}_2$, in contrast, different forms of the dipole operator will mainly lead to a different weighting of the contributions from individual orbitals to the overall corrections. For instance, in the length-form case, the contributions of the $2\sigma_u$ orbital are prevalent, while if the velocity form is taken, the contributions from both $\sigma_u$ orbitals are comparable. Another interesting feature is that both forms lead to a sharp harmonic suppression near $\theta_L=(2n+1)\pi/2$. This is due to the fact that the nodal planes of the $1\sigma_u$ and $2\sigma_u$ orbitals are parallel to the laser-field polarization for such angles.

In CO, however, we notice two interesting features. Firstly, the nodes disappear if the length form is taken. This is partly due to the static dipole moment, which is only present in the length form and leads to corrections which behave like the HOMO. This implies that the overall dipole will decrease for $\theta_L=\pi$ as shown in the figure. In contrast, in the velocity form, as this term is absent, there are minima near $\theta_L=\pi/4$. These minima come from the geometry of  $2\sigma$ and $4\sigma$, which are the heteronuclear counterparts of $1\sigma_u$ and $2\sigma_u$, leading to a strong suppression in the yield close to $\theta_L=(2n+1)\pi/2$. They are, however, slightly shifted from these values due to the heteronuclear character of the molecule. The remaining $\sigma$ orbitals shift these minima even further, to $\theta_L \simeq \pi/4$.

\section{Conclusions}

\label{concl}

In conclusion, our
results show that an adequate choice of the dipole operator in order to
model high-order harmonic generation is still an open question. In the
single-active-electron approximation, enough evidence has been provided in
\cite{JMOCL2007}  that the velocity form yields the best agreement with the
double-slit physical picture. If one goes beyond
this approximation, however, the velocity form leads to a vanishing static
dipole moment for heteronuclear molecules. This may be a particularity of
the basis set involved.

Another issue in the velocity form are the so-called overlap
integrals, in which the dipole operator couples wavefunctions
localized at different ions in the molecule. Throughout the
literature, it has been argued that such integrals are small and may
be neglected without loss of information. Our results, however,
indicate that, depending on the basis set used, their outcome may be
comparable to that of the direct integrals, in which wavefunctions
at the same ion are coupled. As the internuclear distance increases,
however, the overlap integrals become vanishing small.

In contrast, the length-form of the dipole operator leads to a
non-vanishing static dipole moment for a heteronuclear molecule, and
very small overlap integrals, in comparison to their direct
counterparts. The main problem, however, is that such a formulation
leads to terms in the dipole-matrix elements with increase
indefinitely with the internuclear separation. They also couple
states with the same parity, which is unphysical for an odd operator
such as the dipole. Such terms have been also found in the
single-active electron context \cite{JMOCL2007}. Hence, it seems
that the velocity form still gives better results.

Furthermore, suppression of the harmonic spectra caused by the presence of nodes in the orbitals is far more robust than the maxima and minima coming from two-center interference. Interestingly, for heteronuclear molecules, despite interference minima being washed out, the high-harmonic suppression caused by the geometry of the orbitals is still present in the spectra.  However, compared to their homonuclear counterpart, this suppression occurs at different alignment angles.

Finally, in order to incorporate
multielectron effects in SFA-like models in an appropriate way, one
must consider the dynamics of the target. Concrete examples are excitation,
relaxation, electron orbits starting or finishing at different molecular
orbitals, or the motion of the bound electrons. This paper has shown that, if
multielectron effects are considered statically by employing many-body
perturbation theory around the single-electron strong-field
approximation, they lead to corrections which are several
orders of magnitude smaller than the zeroth-order spectrum. This holds for
both homonuclear and
heteronuclear molecules. In fact, recent results in which excitation or
relaxation have been incorporated numerically exhibited larger
corrections \cite{Smirnova,Patchkovskii}.

\textbf{Acknowledgements:} We profited from several discussions with H. J. J. van Dam, P. J. Durham, P. Sherwood and in particular J. Tennyson. This work has been funded by the UK EPSRC and by the Daresbury Laboratory.
\section{Appendix 1: Overlap integrals for dipole matrix elements}

In this appendix, we are providing the generalized expressions for the
overlap integrals (\ref{overlap}) considering the dipole operator in its
velocity and length forms.

\subsection{Velocity form}

We will commence by the velocity-form expressions. If only $\sigma $
orbitals are coupled,one may show that
\begin{eqnarray}  \label{staticdip}
\mathcal{I}_{\xi ,\xi ^{\prime }}^{\alpha ,\beta }(\sigma ,\sigma ) &=&%
\hspace*{-0.2cm}\sum_{j,j^{\prime }}\frac{\pi b_{j,\nu }^{(\xi
)}b_{j^{\prime },\mu }^{(\xi ^{\prime })}}{\zeta _{j,\nu }^{(\xi )}+\zeta
_{j^{\prime },\mu }^{(\xi ^{\prime })}}e^{-\frac{\zeta _{j,\nu }^{(\xi
)}\zeta _{j^{\prime },\mu }^{(\xi ^{\prime })}}{\zeta _{j,\nu }^{(\xi
)}+\zeta _{j^{\prime },\mu }^{(\xi ^{\prime })}}R^{2}} \\
&&\times \lbrack l_{\beta }\mathcal{J}(l_{\alpha },l_{\beta }-1)-2\zeta
_{j^{\prime },\mu }^{(\xi ^{\prime })}\mathcal{J}(l_{\alpha },l_{\beta }+1)],
\notag  \label{explicitoverlap}
\end{eqnarray}%
where
\begin{eqnarray}
\mathcal{J}(l_{\alpha },l_{\beta }) &=&\int_{-\infty }^{\infty }\left[ u\pm
\frac{\rho _{j^{\prime },j,\mu ,\nu }^{(\xi )}}{2}R\right] ^{l_{\beta }}%
\left[ u\mp \frac{\rho _{j,j^{\prime },\nu ,\mu }^{(\xi ^{\prime })}}{2}R%
\right] ^{l\alpha }  \notag \\
&&\times \exp [-(\zeta _{j,\nu }^{(\xi )}+\zeta _{j^{\prime },\mu }^{(\xi
^{\prime })})u^{2}]du,
\end{eqnarray}%
with%
\begin{equation}
\rho _{j,j^{\prime },\nu ,\mu }^{(\xi )}=\frac{-2\zeta _{j^{\prime },\mu
}^{(\xi )}}{\zeta _{j,\nu }^{(\xi )}+\zeta _{j^{\prime },\mu }^{(\xi )}},
\label{rho}
\end{equation}%
\begin{equation}
u=z\pm \frac{\zeta _{j,\nu }^{(\xi )}-\zeta _{j^{\prime },\mu }^{(\xi
^{\prime })}}{2(\zeta _{j,\nu }^{(\xi )}+\zeta _{j^{\prime },\mu }^{(\xi
^{\prime })})}R
\end{equation}%
and $\mathcal{J}(l_{\alpha },l_{\beta }-1)=0$ for $l_{\beta }=0.$

If, on the other hand, we consider a transition from a $\sigma $ to a $\pi
_{x}$ or $\pi _{y}$ orbital,
\begin{widetext}
\begin{eqnarray}
\mathcal{I}_{\xi ,\xi ^{\prime }}^{\alpha ,\beta }(\sigma ,\pi _{\chi }) &=&%
\hspace*{-0.2cm}\sum_{j,j^{\prime }}\frac{\sqrt{\pi }b_{j,\nu }^{(\xi
)}b_{j^{\prime },\mu }^{(\xi ^{\prime })}}{\left( \zeta _{j,\nu }^{(\xi
)}+\zeta _{j^{\prime },\mu }^{(\xi ^{\prime })}\right) ^{1/2}}e^{-\frac{%
\zeta _{j,\nu }^{(\xi )}\zeta _{j^{\prime },\mu }^{(\xi ^{\prime })}}{\zeta
_{j,\nu }^{(\xi )}+\zeta _{j^{\prime },\mu }^{(\xi ^{\prime })}}R^{2}} \notag\\
&&\times \left[ l_{\beta }\mathcal{C}(l_{\beta }-1,l_{\alpha
},0)-2\zeta _{j^{\prime },\mu }^{(\xi ^{\prime })}\left[
\mathcal{C}(l_{\beta },l_{\alpha },1)+\mathcal{C}(l_{\beta
}+1,l_{\alpha },0)\right] \right] ,
\end{eqnarray}%
where $\mathcal{C}(l_{\alpha },l_{\beta },l_{\gamma
})=\mathcal{F}(l_{\alpha })\mathcal{J}(l_{\beta },l_{\gamma }).$ The
function $\mathcal{F}(l_{\alpha })$ is defined in Eq. (\ref{F}). If
only $\pi _{x}$ or $\pi _{y}$ orbitals
are coupled, the overlap integrals read%
\begin{eqnarray}
\mathcal{I}_{\xi ,\xi ^{\prime }}^{\alpha ,\beta }(\pi _{\chi },\pi _{\chi
}) &=&\hspace*{-0.2cm}\sum_{j,j^{\prime }}\frac{\sqrt{\pi }b_{j,\nu }^{(\xi
)}b_{j^{\prime },\mu }^{(\xi ^{\prime })}}{\left( \zeta _{j,\nu }^{(\xi
)}+\zeta _{j^{\prime },\mu }^{(\xi ^{\prime })}\right) ^{1/2}}e^{-\frac{%
\zeta _{j,\nu }^{(\xi )}\zeta _{j^{\prime },\mu }^{(\xi ^{\prime })}}{\zeta
_{j,\nu }^{(\xi )}+\zeta _{j^{\prime },\mu }^{(\xi ^{\prime })}}R^{2}}
\notag \\
&\times &\left[ l_{\beta }\mathcal{C}(l_{\beta }+l_{\alpha }-1,0,0)-2\zeta
_{j^{\prime },\mu }^{(\xi ^{\prime })}\left[ \mathcal{C}(l_{\beta
}+l_{\alpha },0,1)+\mathcal{C}(l_{\beta }+l_{\alpha }+1,0,0)\right] \right] ,
\end{eqnarray}%
\end{widetext}with $\chi =x$ or $y.$ Finally, if we consider the overlap
integral coupling different $\pi $ orbitals, we will find%
\begin{eqnarray}
\mathcal{I}_{\xi ,\xi ^{\prime }}^{\alpha ,\beta }(\pi _{\lambda },\pi
_{\chi }) &=&\sum_{j,j^{\prime }}\frac{\sqrt{\pi }b_{j,\nu }^{(\xi
)}b_{j^{\prime },\mu }^{(\xi ^{\prime })}}{\left( \zeta _{j,\nu }^{(\xi
)}+\zeta _{j^{\prime },\mu }^{(\xi ^{\prime })}\right) ^{1/2}}e^{-\frac{%
\zeta _{j,\nu }^{(\xi )}\zeta _{j^{\prime },\mu }^{(\xi ^{\prime })}}{\zeta
_{j,\nu }^{(\xi )}+\zeta _{j^{\prime },\mu }^{(\xi ^{\prime })}}R^{2}} \\
&&\times \{-2\zeta _{j^{\prime },\mu }^{(\xi ^{\prime })}\mathcal{G}%
(l_{\alpha },l_{\beta })+l_{\beta }\mathcal{H}(l_{\alpha },l_{\beta
}-1,0,0)\}.  \notag  \label{overlaplength}
\end{eqnarray}%
with
\begin{eqnarray}
\mathcal{G}(l_{\alpha },l_{\beta }) &=&\mathcal{H}(l_{\alpha },l_{\beta
},0,1) \\
&&+\mathcal{H}(l_{\alpha }+1,l_{\beta },0,0)+\mathcal{H}(l_{\alpha
},l_{\beta }+1,0,0)  \notag
\end{eqnarray}%
and $\mathcal{H}(l_{\alpha },l_{\beta },l_{\gamma },l_{\delta })=\mathcal{A}%
(l_{\alpha },l_{\beta })\mathcal{J}(l_{\gamma },l_{\delta }).$ In Eq. (\ref%
{overlaplength}), $\lambda \neq \chi .$ The function $\mathcal{A}(l_{\alpha
},l_{\beta })$ is defined in Eq. (\ref{pisigm1}).

\subsection{Length form}

For the length form of the dipole operator, if only transitions involving $%
\sigma $ orbitals are considered, we find that%
\begin{equation}
\mathcal{I}_{\xi ,\xi ^{\prime }}^{\alpha ,\beta }(\sigma ,\sigma )=\hspace*{%
-0.2cm}\sum_{j,j^{\prime }}\frac{\pi b_{j,\nu }^{(\xi )}b_{j^{\prime },\mu
}^{(\xi ^{\prime })}}{\zeta _{j,\nu }^{(\xi )}+\zeta _{j^{\prime },\mu
}^{(\xi ^{\prime })}}e^{-\frac{\zeta _{j,\nu }^{(\xi )}\zeta _{j^{\prime
},\mu }^{(\xi ^{\prime })}}{\zeta _{j,\nu }^{(\xi )}+\zeta _{j^{\prime },\mu
}^{(\xi ^{\prime })}}R^{2}}\mathcal{K}(l_{\alpha },l_{\beta }),
\end{equation}%
where
\begin{eqnarray}
\mathcal{K}(l_{\alpha },l_{\beta }) \hspace*{-0.15cm}&\hspace*{-0.1cm}&=%
\hspace*{-0.15cm}\int_{-\infty }^{\infty }\left[ u\pm \frac{\rho _{j^{\prime
},j,\mu ,\nu }^{(\xi )}}{2}R\right] ^{l_{\beta }}\hspace*{-0.15cm}\left[
u\mp \frac{\rho _{j,j^{\prime },\nu ,\mu }^{(\xi ^{\prime })}}{2}R\right]
^{l\alpha } \\
\hspace*{-0.15cm}&\hspace*{-0.15cm}&\hspace*{-0.15cm}\times \left[ u\mp
\frac{\zeta _{j,\nu }^{(\xi )}-\zeta _{j^{\prime },\mu }^{(\xi ^{\prime })}}{%
2(\zeta _{j,\nu }^{(\xi )}+\zeta _{j^{\prime },\mu }^{(\xi ^{\prime })})}R%
\right] \exp [-(\zeta _{j,\nu }^{(\xi )}+\zeta _{j^{\prime },\mu }^{(\xi
^{\prime })})u^{2}]du.  \notag
\end{eqnarray}%
For transitions from a $\sigma $ to a $\pi _{x}$ or $\pi _{y}$ orbital, the
overlap integral reads%
\begin{eqnarray}
\mathcal{I}_{\xi ,\xi ^{\prime }}^{\alpha ,\beta }(\sigma ,\pi _{\chi }) &=&%
\hspace*{-0.2cm}\sum_{j,j^{\prime }}\frac{\sqrt{\pi }b_{j,\nu }^{(\xi
)}b_{j^{\prime },\mu }^{(\xi ^{\prime })}}{\left( \zeta _{j,\nu }^{(\xi
)}+\zeta _{j^{\prime },\mu }^{(\xi ^{\prime })}\right) ^{1/2}}e^{-\frac{%
\zeta _{j,\nu }^{(\xi )}\zeta _{j^{\prime },\mu }^{(\xi ^{\prime })}}{\zeta
_{j,\nu }^{(\xi )}+\zeta _{j^{\prime },\mu }^{(\xi ^{\prime })}}R^{2}} \\
&&\times \left\{ \mathcal{F}[l_{\alpha }+1]\mathcal{J}(0,l_{\beta })+%
\mathcal{F}[l_{\alpha }]\mathcal{K}(0,l_{\beta })\right\} ,  \notag
\end{eqnarray}%
with $\chi =x$ or $y.$

Finally, if only $\pi $ orbitals are involved, one may distinguish two types
of overlap integrals. Either the dipole matrix element couples the same
initial and final orbital, $\pi _{x}$ or $\pi _{y},$or the orbitals $\pi
_{\chi }$ and $\pi _{\lambda },$ $\chi \neq \lambda .$ The former case is
given by
\begin{eqnarray}
\mathcal{I}_{\xi ,\xi ^{\prime }}^{\alpha ,\beta }(\pi _{\chi },\pi _{\chi })%
\hspace*{-0.2cm} &&\hspace*{-0.2cm}=\hspace*{-0.15cm}\sum_{j,j^{\prime }}%
\frac{\sqrt{\pi }b_{j,\nu }^{(\xi )}b_{j^{\prime },\mu }^{(\xi ^{\prime })}}{%
\left( \zeta _{j,\nu }^{(\xi )}+\zeta _{j^{\prime },\mu }^{(\xi ^{\prime
})}\right) ^{1/2}}e^{-\frac{\zeta _{j,\nu }^{(\xi )}\zeta _{j^{\prime },\mu
}^{(\xi ^{\prime })}}{\zeta _{j,\nu }^{(\xi )}+\zeta _{j^{\prime },\mu
}^{(\xi ^{\prime })}}R^{2}} \\
&&\hspace*{-0.2cm}\times \left\{ \mathcal{F}[l_{\alpha }+l_{\beta }+1]%
\mathcal{J}(0,0)+\mathcal{F}[l_{\alpha }+l_{\beta }]\mathcal{K}(0,0)\right\}
,  \notag
\end{eqnarray}%
while the latter reads%
\begin{eqnarray}
\mathcal{I}_{\xi ,\xi ^{\prime }}^{\alpha ,\beta }(\pi _{\lambda },\pi
_{\chi }) &=&\hspace*{-0.2cm}\sum_{j,j^{\prime }}\frac{\sqrt{\pi }b_{j,\nu
}^{(\xi )}b_{j^{\prime },\mu }^{(\xi ^{\prime })}}{\left( \zeta _{j,\nu
}^{(\xi )}+\zeta _{j^{\prime },\mu }^{(\xi ^{\prime })}\right) ^{1/2}}e^{-%
\frac{\zeta _{j,\nu }^{(\xi )}\zeta _{j^{\prime },\mu }^{(\xi ^{\prime })}}{%
\zeta _{j,\nu }^{(\xi )}+\zeta _{j^{\prime },\mu }^{(\xi ^{\prime })}}R^{2}}
\\
&&\times \left\{ \mathcal{E}(l_{\alpha },l_{\beta })\mathcal{J}(0,0)+%
\mathcal{A}(l_{\alpha },l_{\beta })\mathcal{K}(0,0)\right\} ,  \notag
\end{eqnarray}%
with $\mathcal{E}(l_{\alpha },l_{\beta })=$ $\mathcal{A}(l_{\alpha
},l_{\beta }+1)+\mathcal{A}(l_{\alpha }+1,l_{\beta }).$

\subsection{Particular integrals $\mathcal{J}$ and $\mathcal{K}$}

Specifically for the basis set employed in this work, we find%
\begin{equation}
\mathcal{J}(0,0)=\sqrt{\frac{\pi }{\zeta _{j,\nu }^{(\xi )}+\zeta
_{j^{\prime },\mu }^{(\xi ^{\prime })}}},
\end{equation}

\begin{equation}
\mathcal{J}(0,1)=\pm \frac{\sqrt{\pi }\zeta _{j,\nu }^{(\xi )}R}{\left(
\zeta _{j,\nu }^{(\xi )}+\zeta _{j^{\prime },\mu }^{(\xi ^{\prime })}\right)
^{3/2}},
\end{equation}

\begin{equation}
\mathcal{J}(1,1)=\frac{\sqrt{\pi }}{2\left( \zeta _{j,\nu }^{(\xi )}+\zeta
_{j^{\prime },\mu }^{(\xi ^{\prime })}\right) ^{3/2}}\left[ 1-\frac{2\zeta
_{j,\nu }^{(\xi )}\zeta _{j^{\prime },\mu }^{(\xi ^{\prime })}}{\zeta
_{j,\nu }^{(\xi )}+\zeta _{j^{\prime },\mu }^{(\xi ^{\prime })}}R^{2}\right]
,
\end{equation}%
\begin{equation}
\mathcal{J}(0,2)=\frac{\sqrt{\pi }}{2\left( \zeta _{j,\nu }^{(\xi )}+\zeta
_{j^{\prime },\mu }^{(\xi ^{\prime })}\right) ^{3/2}}\left[ 1+\frac{2\left(
\zeta _{j,\nu }^{(\xi )}\right) ^{2}}{\zeta _{j,\nu }^{(\xi )}+\zeta
_{j^{\prime },\mu }^{(\xi ^{\prime })}}R^{2}\right]
\end{equation}%
and%
\begin{eqnarray}
\mathcal{J}(1,2) &=&\pm \frac{\sqrt{\pi }}{2\left( \zeta _{j,\nu }^{(\xi
)}+\zeta _{j^{\prime },\mu }^{(\xi ^{\prime })}\right) ^{5/2}} \\
&&\times \left[ (\zeta _{j^{\prime },\mu }^{(\xi ^{\prime })}-2\zeta _{j,\nu
}^{(\xi )})R+\frac{2\zeta _{j^{\prime },\mu }^{(\xi ^{\prime })}\left( \zeta
_{j,\nu }^{(\xi )}\right) ^{2}}{\left( \zeta _{j,\nu }^{(\xi )}+\zeta
_{j^{\prime },\mu }^{(\xi ^{\prime })}\right) }R^{3}\right] .  \notag
\end{eqnarray}%
for the integrals $\mathcal{J}(l_{\alpha },l_{\beta }\pm 1)$ and $\mathcal{J}%
(0,l_{\beta }),$ and
\begin{equation}
\mathcal{K}(0,0)=\mp \sqrt{\frac{\pi }{\zeta _{j,\nu }^{(\xi )}+\zeta
_{j^{\prime },\mu }^{(\xi ^{\prime })}}}\frac{\zeta _{j,\nu }^{(\xi )}-\zeta
_{j^{\prime },\mu }^{(\xi ^{\prime })}}{2(\zeta _{j,\nu }^{(\xi )}+\zeta
_{j^{\prime },\mu }^{(\xi ^{\prime })})}R,
\end{equation}

\begin{equation}
\mathcal{K}(0,1)=\frac{\sqrt{\pi }}{2\left( \zeta _{j,\nu }^{(\xi )}+\zeta
_{j^{\prime },\mu }^{(\xi ^{\prime })}\right) ^{3/2}}\left[ 1+ \frac{\zeta
_{j,\nu }^{(\xi )}(\zeta _{j,\nu }^{(\xi )}-\zeta _{j^{\prime },\mu }^{(\xi
^{\prime })})}{2\left( \zeta _{j,\nu }^{(\xi )}+\zeta _{j^{\prime },\mu
}^{(\xi ^{\prime })}\right) }R^{2}\right] ,
\end{equation}%
\begin{equation}
\mathcal{K}(1,0)=\frac{\sqrt{\pi }}{2\left( \zeta _{j,\nu }^{(\xi )}+\zeta
_{j^{\prime },\mu }^{(\xi ^{\prime })}\right) ^{3/2}}\left[ 1- \frac{\zeta
_{j^{\prime },\mu }^{(\xi ^{\prime })}(\zeta _{j,\nu }^{(\xi )}-\zeta
_{j^{\prime },\mu }^{(\xi ^{\prime })})}{2\left( \zeta _{j,\nu }^{(\xi
)}+\zeta _{j^{\prime },\mu }^{(\xi ^{\prime })}\right) }R^{2}\right]
\end{equation}%
and
\begin{equation}
\mathcal{K}(1,1)=\mp \frac{\sqrt{\pi }(\zeta _{j,\nu }^{(\xi )}-\zeta
_{j^{\prime },\mu }^{(\xi ^{\prime })})}{2\left( \zeta _{j,\nu }^{(\xi
)}+\zeta _{j^{\prime },\mu }^{(\xi ^{\prime })}\right) ^{5/2}}\left[ \frac{3R%
}{2}-\frac{\zeta _{j,\nu }^{(\xi )}\zeta _{j^{\prime },\mu }^{(\xi ^{\prime
})}}{\zeta _{j,\nu }^{(\xi )}+\zeta _{j^{\prime },\mu }^{(\xi ^{\prime })}}%
R^{3}\right]
\end{equation}%
for the integrals $\mathcal{K}(l_{\alpha },l_{\beta })$. The above-stated
expressions show that the integrals $\mathcal{J}(l_{\alpha },l_{\beta }\pm
1) $ and $\mathcal{K}(l_{\alpha },l_{\beta })$ will lead to a polynomial
dependence on $R/(\zeta _{j,\nu }^{(\xi )}+\zeta _{j^{\prime },\mu }^{(\xi
^{\prime })})^{1/2+n}$. This dependence will be damped by the gaussian
factor in Eq. (\ref{explicitoverlap}), so that the overlap integrals are
expected to decrease with the internuclear distance.

\section{Appendix 2: Angle-integrated dipole matrix elements}

In this appendix, we provide the expressions obtained for the dipole matrix
elements, if the azimuthal angular variable is integrated over. Let us start
with $\mathrm{N}_{2}$. In this case, the HOMO is a $3\sigma _{g} $ orbital.
The orbitals that will contribute to the corrections are $1\pi _{u_{x}}$, $%
1\pi _{u_{y}}$, $2\sigma _{u}$ and 1$\sigma _{u}$. The remaining, $\sigma
_{g}$ orbitals do not contribute to the corrections for symmetry reasons. We
will be able then to write for the first-order dipole matrix element,

\begin{equation}
d^{(1)}(\mathbf{k}+\mathbf{A}(\tau ))=d_{3\sigma _{g}}^{(0)}(\mathbf{k}+%
\mathbf{A}(\tau ))+d_{\sigma \sigma }^{(1)}+d_{\sigma \pi
_{x}}^{(1)}+d_{\sigma \pi _{y}}^{(1)},  \label{1storderN2}
\end{equation}%
where
\begin{equation}
d_{\sigma \sigma }^{(1)}=-\sum^2_{n=1}d_{3\sigma _{g},n\sigma _{u}}\psi
_{n\sigma _{u}}(\mathbf{k}+\mathbf{A}(\tau ))
\end{equation}
and
\begin{equation}
d_{\sigma \pi _{\chi }}^{(1)}=-d_{3\sigma _{g},1\pi _{u\chi }}\psi _{1\pi
_{u\chi }}(\mathbf{k}+\mathbf{A}(\tau )).  \label{dipolesigmapi}
\end{equation}%
In the above-stated equations, $\tau =t,t^{\prime }$ and $\chi =x,y.$ With
respect to the dependence on the azimuthal coordinate $\phi _{k}$, Eq. (\ref%
{1storderN2}) can be written as
\begin{eqnarray}
d^{(1)}(\mathbf{k}+\mathbf{A}(\tau )) &=&\mathcal{D}_{\sigma }(k,\tau
,\theta _{k})+\mathcal{D}_{\pi _{x}}(k,\tau ,\theta _{k})\cos [\phi _{k}]
\notag \\
&&+\mathcal{D}_{\pi _{y}}(k,\tau ,\theta _{k})\sin [\phi _{k}],
\end{eqnarray}%
where $\mathcal{D}(k,\tau ,\theta _{k})$ indicates the part of the
prefactors which do not depend on the azimuthal coordinate $\phi _{k}$. The
modulus square of the product $\left[ d^{(1)}(\mathbf{k}+\mathbf{A}(t))%
\right] ^{\ast }d^{(1)}(\mathbf{k}+\mathbf{A}(t^{\prime }),$ when integrated
over the azimuthal coordinate, leads to the effective prefactor
\begin{eqnarray}
\mathcal{\tilde{D}}(k,t,t^{\prime },\theta _{k})) &=&2\pi \Delta
_{1}(k,t,t^{\prime },\theta _{k})+\pi \Delta _{2}(k,t,t^{\prime },\theta
_{k}) \\
&&+\frac{3\pi }{4}\Delta _{3}(k,t,t^{\prime },\theta _{k})+\frac{\pi }{4}%
\Delta _{4}(k,t,t^{\prime },\theta _{k}),  \notag
\end{eqnarray}%
where the constants arise from integrals of the form $\int \sin ^{m}[\phi
_{k}]\cos ^{n}[\phi _{k}]d\phi _{k},$ where $m,n$ are even integer numbers.
The explicit expressions for the functions $\Delta _{n}(k,t,t^{\prime
},\theta _{k})$ read
\begin{equation}
\Delta _{1}(k,t,t^{\prime },\theta _{k})=\left\vert \mathcal{D}_{\sigma
\sigma }(k,t,t^{\prime },\theta _{k})\right\vert ^{2},
\end{equation}
\begin{widetext}
\begin{eqnarray}
\Delta _{2}(k,t,t^{\prime },\theta _{k}) &=&\left\vert \mathcal{D}_{\sigma
\pi _{x}}(k,t,t^{\prime },\theta _{k})\right\vert ^{2}+\left\vert \mathcal{D}%
_{\sigma \pi _{y}}(k,t,t^{\prime },\theta _{k})\right\vert ^{2}+\mathcal{D}%
_{\sigma \sigma }(k,t,t^{\prime },\theta _{k})(\mathcal{D}_{\pi _{x}\pi
_{x}}^{\ast }(k,t,t^{\prime },\theta _{k})+\mathcal{D}_{\pi _{y}\pi
_{y}}^{\ast }(k,t,t^{\prime },\theta _{k}))  \notag \\
&&+\mathcal{D}_{\sigma \sigma }^{\ast }(k,t,t^{\prime },\theta _{k})(%
\mathcal{D}_{\pi _{x}\pi _{x}}(k,t,t^{\prime },\theta _{k})+\mathcal{D}_{\pi
_{y}\pi _{y}}(k,t,t^{\prime },\theta _{k})),
\end{eqnarray}%
\begin{equation}
\Delta _{3}(k,t,t^{\prime },\theta _{k})=\left\vert \mathcal{D}_{\pi _{x}\pi
_{x}}(k,t,t^{\prime },\theta _{k})\right\vert ^{2}+\left\vert \mathcal{D}%
_{\pi _{y}\pi _{y}}(k,t,t^{\prime },\theta _{k})\right\vert ^{2}
\end{equation}%
\begin{equation}
\Delta _{4}(k,t,t^{\prime },\theta _{k})=\left\vert \mathcal{D}_{\pi _{x}\pi
_{y}}(k,t,t^{\prime },\theta _{k})\right\vert ^{2}+\mathcal{D}_{\pi _{x}\pi
_{x}}^{\ast }(k,t,t^{\prime },\theta _{k})\mathcal{D}_{\pi _{y}\pi
_{y}}(k,t,t^{\prime },\theta _{k})+\mathcal{D}_{\pi _{x}\pi
_{x}}(k,t,t^{\prime },\theta _{k})\mathcal{D}_{\pi _{y}\pi _{y}}^{\ast
}(k,t,t^{\prime },\theta _{k}),  \notag
\end{equation}%
and where%
\begin{equation}
\mathcal{D}_{ij}(k,t,t^{\prime },\theta _{k})=\left\{
\begin{array}{l}
\mathcal{D}_{i}^{\ast }(k,t,\theta _{k})\mathcal{D}%
_{j}(k,t^{\prime },\theta _{k}),i=j \\
\mathcal{D}_{i}^{\ast }(k,t,\theta _{k})\mathcal{D}_{j}(k,t^{\prime },\theta
_{k})+\mathcal{D}_{j}^{\ast }(k,t,\theta _{k})\mathcal{D}_{i}(k,t^{\prime
},\theta _{k}),i\neq j.%
\end{array}%
\right.
\end{equation}
\end{widetext}
For CO, there will be two main differences. Firstly, a static dipole moment
is expected. Secondly, due to the fact that the orbitals do not exhibit a
well-defined parity, more states are coupled by the dipole operator through
the matrix elements $d_{0i}$. In fact, apart from the HOMO, there will now
be four contributing occupied $\sigma$ states and the two degenerate $\pi$
states corresponding to the HOMO-1. The explicit expressions for $%
d^{(1)}_{\sigma\sigma}$ will change to
\begin{equation}
d^{(1)}_{\sigma\sigma}=-\sum^4_{n=1}d_{5\sigma,n\sigma }\psi _{n\sigma}(%
\mathbf{k}+\mathbf{A}(\tau ))+d^{(1)}_{\mathrm{static}},
\end{equation}
with
\begin{equation}
d^{(1)}_{\mathrm{static}}=\sum^4_{n=1}d_{n\sigma,n\sigma }\psi _{5\sigma}(%
\mathbf{k}+\mathbf{A}(\tau )).
\end{equation}
This will lead to changes in $\mathcal{D}_{\sigma }(k,\tau,\theta_k)$.
Furthermore, in Eq.~(\ref{dipolesigmapi}) $3\sigma_g$ should now be replaced
by $5\sigma$ and $1\pi_{u\chi}$ by $1\pi_{\chi}$. There will be, however, no
change in the structure of the subsequent equations or in the weights
obtained upon angular integration.

\end{document}